\documentstyle[preprint,aps,floats,epsfig,subeqn,amssymb]{revtex}

\newcommand{\be}{\begin{equation}}
\newcommand{\ee}{\end{equation}}
\newcommand{\bea}{\begin{eqnarray}}
\newcommand{\eea}{\end{eqnarray}}
\newcommand{\bml}{\begin{mathletters}}
\newcommand{\eml}{\end{mathletters}}

\newcounter{fixy}
\begin{document}
\newenvironment{fixy}[1]{\setcounter{figure}{#1}}
{\addtocounter{fixy}{1}}
\renewcommand{\thefixy}{\arabic{fixy}}
\renewcommand{\thefigure}{\thefixy\alph{figure}}
\setcounter{fixy}{1}
\tighten

\preprint{DCPT-02/75}
\draft




\title{SU(5) monopoles and non-abelian black holes}
\renewcommand{\thefootnote}{\fnsymbol{footnote}}
\author{ Yves Brihaye\footnote{Yves.Brihaye@umh.ac.be}}
\address{Facult\'e des Sciences, Universit\'e de Mons-Hainaut,
 B-7000 Mons, Belgium}
\author{Betti Hartmann\footnote{Betti.Hartmann@durham.ac.uk}}
\address{Department of Mathematical Sciences, University
of Durham, Durham DH1 3LE, U.K.}
\date{\today}
\setlength{\footnotesep}{0.5\footnotesep}

\maketitle
\begin{abstract}
We construct spherically and axially symmetric monopoles 
in SU(5) Yang-Mills-Higgs theory both in flat
and curved space as well as spherical and axial
non-abelian, ``hairy'' black holes.
We find that in analogy to the SU(2) case, the flat 
space monopoles are either non-interacting (in the BPS limit) or 
repelling. In curved space, however, gravity is able to overcome
the repulsion for suitable choices of the Higgs coupling constants and
the gravitational coupling. In addition, we confirm that indeed all qualitative
features of (gravitating) SU(2) monopoles are found as well in the SU(5) case. 
For the non-abelian black holes, we compare the behaviour of the solutions
in the BPS limit with that for non-vanishing Higgs self-coupling constants.

\end{abstract}


\renewcommand{\thefootnote}{\arabic{footnote}}
\section{Introduction}
Grand Unified Theories (GUTs) are believed to be valid for energies 
above $10^{14}$ GeV (which corresponds to $t=10^{-34}sec$ after the Big Bang)
and unify all known interactions except for gravity. In a number of spontaneous
breakdowns of the symmetry of the GUT, the today present 
SU(3)$_c$ $\times$ SU(2)$_{L}$ $\times$ U(1)$_Y$
symmetry of the universe is attained. For some time, it was believed that the GUT has
a single gauge group, namely SU(5) \cite{gg}, however, since this theory predicts a lifetime
of the proton of about $10^{28}$ to $10^{30}$ years \cite{kt}, while
experimentally it was found to be roughly $10^{33}$ years \cite{ex}, it was soon
dropped as a candidate for a GUT.

The bosonic part of the Georgi-Glashow model with gauge group SU(2) is believed to be
a good toy model for GUTs. It consists of a Higgs field in the adjoint representation
of the gauge group and through the spontaneous symmetry breakdown 
of SU(2) to U(1) two of the three gauge bosons $W_{\pm}$ as well as the Higgs boson
itself gain mass. The massless gauge boson is associated with the unbroken U(1)
symmetry and identified with the photon.

In 1974, 't Hooft and Polyakov \cite{hp} 
made the interesting observation that the bosonic part of the 
Georgi-Glashow model allows for soliton solutions, i.e. particle-like, 
finite energy
solutions, which due to their topological properties carry a non-trivial
magnetic charge and have thus been named "magnetic monopoles".

Since the $n=1$ solution was shown to be the unique spherically symmetric solution
\cite{bogo}, construction of higher winding number solutions longed for
an Ansatz with less symmetry. After their existence has been proved 
\cite{taubes}, axially symmetric monopoles
have been constructed \cite{rr}. These monopoles can be thought of 
as two monopoles superposed on each other at the origin with torus-like
energy density. Since
monopoles in the Bogomol'nyi-Prasad-Sommerfield (BPS) limit
\cite{bogo,ps} are non-interacting \cite{manton,oyp}, while for non-vanishing Higgs
boson mass a Coulomb-like repulsive force acts between them \cite{oyp},
the mass per winding number of the $n$-multimonopole is equal to (resp. bigger
than) $n$ times the mass of the $n=1$ monopole for vanishing (resp. non-vanishing)
Higgs boson mass \cite{kkt}. This leads to the conclusion that in flat space
no bound multimonopoles are possible. However, the inclusion of gravity \cite{hkk}
and/or 
a dilaton \cite{bh} can render attraction if the Higgs boson mass is small enough. 

A lot of work has been done on the topic of 
embedding SU(2) monopoles into higher gauge groups \cite{su3,nb,tomaras,eweinberg}.
The embedding into SU(5) has been constructed in \cite{tomaras} where through the
introduction of an 24-dimensional Higgs field in the adjoint representation
as well as a 5-dimensional Higgs field in the fundamental representation
the breakdown of SU(5) to SU(3)$\times$ U(1) is achieved. Because of the above arguments
for SU(5) not being a good canditate for a GUT \cite{newdata}, scientists have lost interest in this
model.

Recently, however, it was shown \cite{va} that the monopole spectrum
produced in the breakdown of SU(5) to (SU(3) $\times$ SU(2) $\times$
U(1))/ $\mathbb{Z}_{\rm 6}$ corresponds
to the spectrum of one family of fermions in the standard model (SM). Thus interest 
has grown again. An explicit analytical BPS solution was constructed in
this model \cite{meckes}.

Here we construct axially symmetric SU(5) monopoles both in flat and 
curved space as well as the corresponding black hole solutions.
We give the model and the Ansatz in section II. We discuss the 
flat space solutions in
section III, the gravitating solutions in section IV and the non-abelian
black hole solutions in section V. We give our conclusions in
section VI.

\section{The model}
We consider an SU(5) Einstein-Yang-Mills-Higgs (EYMH) model with the following action:
\begin{equation}
S=\int \sqrt{-g}d^4 x \{\frac{R}{16\pi G} + \cal{L}\rm_M \}
\end{equation}
where the matter Langragian $\cal{L}_M$ is given by
\begin{equation}
\cal{L}\rm_M=-\frac{1}{2} Tr(F_{\mu\nu}F^{\mu\nu})- Tr(D_{\mu}\Phi
D^{\mu}\Phi)-V(\Phi)
\end{equation}
with field strength tensor
\begin{equation}
F_{\mu\nu}=\partial_{\mu} A_{\nu}-\partial_{\nu} A_{\mu}-ie[A_{\mu},A_{\nu}]
\end{equation}
and covariant derivative
\begin{equation}
D_{\mu}\Phi=\partial_{\mu}-ie[A_{\mu},\Phi]
\end{equation}
such that
\begin{equation}
A_{\mu}=\frac{1}{2}A^a_{\mu} t^a \ , \ \Phi_{\mu}=\frac{1}{2}\Phi_{\mu}^a t^a  \ , \ a=1,..,24
\end{equation}
and the $t^a$ fulfill the su(5) Lie-Algebra. $G$ denotes Newton's constant
and $e$ the gauge field coupling.

The potential is given by \cite{va,va2}:
\begin{equation}
V(\Phi)=-\lambda_1 Tr(\Phi^2)+\lambda_2 (Tr(\Phi^2))^2+
\lambda_3 Tr(\Phi^4)-V_{min}
\end{equation}
where we have substracted 
$V_{min}=-15\lambda_1^2/(60\lambda_2+14\lambda_3)$ from the potential
in order to make $V(\Phi)$ vanish (and thus have finite energy solutions) when $\Phi$
attains its vacuum expectation value 
$||\Phi||_0=\eta=(\sqrt{\lambda_1}/\sqrt{60\lambda_2+
14\lambda_3}) \rm{diag}(2,-3,2,2,-3)$.
The vacuum expectation value (vev) leads to a spontaneous breakdown of SU(5)
to (SU(3) $\times$ SU(2) $\times$ U(1))/ $\mathbb{Z}_{\rm 6}$, where
$\mathbb{Z}_{\rm 6}=\mathbb{Z}_{\rm 3}\times \mathbb{Z}_{\rm 2}$
denotes the product of the center $\mathbb{Z}_{\rm 3}$ of SU(3) and the
center $\mathbb{Z}_{\rm 2}$ of SU(2). Note that the solutions have an 
additional $\mathbb{Z}_{\rm 2}$ symmetry because we have left out a possible 
cubic term in the potential.
The reason for this is that the inclusion of a $\Phi^3$-term in the potential
would not lead to an algebraically expressible vev. The vev would have to be computed
numerically.

$12$ of the $24$ gauge fields gain mass $M_W$, while $12$ of the Higgs fields become
massive with $8$ fields obtaining the mass $M_{H_{1}}$, $3$ fields
obtaining the mass $M_{H_{2}}$ and $1$ field obtaining the mass $M_{H_{3}}$ denoting
the $({\bf 8}, {\bf 1})$, $({\bf 3}, {\bf 1})$ and $({\bf 1}, {\bf 1})$, respectively,
representation of SU(3)$\times$SU(2):
\begin{equation}
M_W=\sqrt{\frac{5}{6}}e\eta \ , \ \ M_{H_{1}}=\sqrt{10\lambda_3}\eta \ , \ \
M_{H_{2}}=2M_{H_{1}} \ , \ \ M_{H_{3}}=2\eta\sqrt{30\lambda_2+7\lambda_3} \ .
\label{mass}
\end{equation}
\subsection{The Ansatz}
The Ansatz for the gauge and Higgs fields is chosen such that the SU(2) monopole
is embedded in the SU(5) theory, where the
embedding corresponds to $5\rightarrow 1\oplus 2\oplus 3$.

The Ansatz for the gauge fields reads using the Ansatz for the
SU(2) monopoles \cite{rr}:
\begin{equation}
A_{\mu} dx^{\mu}=\frac{1}{2er}\left(t^n_{\varphi}(H_1 dr + (1-H_2)r d\theta
-n(t_{r}^{n}H_3+t_{\theta}^{n} (1-H_4))r \sin\theta d \varphi\right)
\end{equation}
while for the in the adjoint representation of the SU(5) group given Higgs field it reads
\begin{equation}
\Phi=\frac{1}{2}\left(\Phi_1 t_{r}^{n}+ \Phi_2 t_{\theta}^{n}+
\Phi_3 t^4+\Phi_4 t^5\right) \ .  
\end{equation}
The $t^n_{\varphi}$, $t_{r}^{n}$ and $t_{\theta}^{n}$ denote the vector 
product of
the vector $(t^1,t^2,t^3)$ of the three $5\times 5$ matrices $t^a$ which fulfill the
Lie-Algebra of SU(2):
\begin{equation} 
t^a=\rm{diag}(\sigma_a,0,0,0) \ , \ a=1,2,3
\end{equation}
with the unit vectors:
\begin{eqnarray}
\vec e_r^{\, n}      &=& 
(\sin \theta \cos n \varphi, \sin \theta \sin n \varphi, \cos \theta)
\ , \nonumber \\
\vec e_\theta^{\, n} &=& 
(\cos \theta \cos n \varphi, \cos \theta \sin n \varphi,-\sin \theta)
\ , \nonumber \\
\vec e_\phi^{\, n}   &=& (-\sin n \varphi, \cos n \varphi,0) 
\ , \label{rtp} \end{eqnarray}
and $t^4$ and $t^5$ are given by diagonal, traceless matrices:
\begin{equation}
 t^4=\frac{1}{\sqrt{3}}\rm{diag}(0,0,1,1,-2) \ , \ \
t^5=\frac{1}{\sqrt{15}}\rm{diag}(-3,-3,2,2,2)
\end{equation}
Since SU(5) has 4 diagonal generators, 
one could also think about inserting a  term $\propto \Phi_5 t^6$ with $t^6=\rm{diag}(0,0,1,-1,0)$.
However, inserting this into the potential and looking for minima, it turns
out that the trivial solution $\Phi_5\equiv 0$ is a vacuum solution. Thus,
without loosing generality we set $\Phi_5$ to zero.
$H_1$, $H_2$, $H_3$, $H_4$ as well as $\Phi_1$, $\Phi_2$ $\Phi_3$ and $\Phi_4$
depend only on $r$ and $\theta$.

For the metric, we use an Ansatz in isotropic coordinates:
\begin{equation}
ds^2=-f(r,\theta)dt^2+\frac{m(r,\theta)}{f(r,\theta)}(dr^2+r^2 d\theta^2)
+\frac{l(r,\theta)}{f(r,\theta)}r^2 \sin^2\theta d\varphi^2
\label{metric}
\end{equation}
Introduction of a rescaled radial coordinate $x=e\eta r$ leads to a set of
coupled partial differential equations which depend only on the following 
parameters:
\begin{equation}
\alpha=\sqrt{G} v \ , \ \ \beta_2=\frac{\sqrt{\lambda_2}}{e} \ , \ \ 
\beta_3=\frac{\sqrt{\lambda_3}}{e}
\end{equation} 
\subsection{Boundary conditions}
The requirement of regularity at the origin for the (multi)monopoles
leads to the following boundary conditions (bcs):
\begin{subequations}
\begin{equation}
\partial_r f(r,\theta)|_{r=0}=\partial_r l(r,\theta)|_{r=0} = 
\partial_r m(r,\theta)|_{r=0}
\end{equation}
\begin{equation}
 \Phi_i(r=0,\theta)=\partial_r \Phi_{i+2}(r,\theta)|_{r=0}
=H_k(r=0,\theta)=0 \ , \
H_{k+1}(r=0,\theta)=1 \ , i=1,2 \ , k=1,3 
\end{equation}
\end{subequations}
while for the non-abelian black holes, these are replaced
by the boundary conditions at the regular horizon $r_h$:
\begin{subequations}
\begin{equation}
f(r=r_h,\theta)= l(r=r_h,\theta) = 
 m(r=r_h,\theta)=0
\end{equation}
\begin{equation}
\partial_r\Phi_i(r,\theta)|_{r=r_h}=H_1(r=r_h,\theta)
=\partial_r H_j(r,\theta)|_{r=r_h}=0 \ , \ i=1,2,3,4 \ , \ j=2,3,4
\end{equation}
\end{subequations}
These latter result from the regularity of the solutions at the horizon
and a suitable gauge condition \cite{hkk}.

In order to have asymptotically flat, finite energy solutions the bcs at
infinity $(r=\infty$) read:
\begin{subequations}
\begin{equation}
f(r=\infty,\theta)=m(r=\infty,\theta)=l(r=\infty,\theta)=1 
\ , 
\end{equation}
\begin{equation}
H_i(r=\infty,\theta)=\Phi_2(r=\infty,\theta)=0 \ , \  i=1,2,3,4 \ , \
\end{equation}
\begin{equation}
\label{higgs}
  \Phi_1(r=\infty,\theta)=\frac{1}{2}\sqrt{\frac{5}{3}} \ , \ 
\Phi_3(r=\infty,\theta)=\frac{\sqrt{5}}{3} \ , \
 \Phi_4(r=\infty,\theta)=\frac{1}{6} \ .
\end{equation}
\end{subequations}
To obtain the right symmetry of the solutions, we set on both the $z$- 
as well as on the $\rho$-axis ($\theta_0=0$ and $\theta_0=\pi/2$, respectively):
\begin{subequations}
\begin{equation}
\partial_{\theta}f(r,\theta)|_{\theta=\theta_0}=
\partial_{\theta}m|_{\theta=\theta_0}=\partial_{\theta}l|_{\theta=\theta_0}=0
\end{equation}
\begin{equation}
H_i(r,\theta=\theta_0)=\Phi_2(r,\theta=\theta_0)=
\partial_{\theta}H_{i+1}(r,\theta)|_{\theta=\theta_0}
=\partial_{\theta}\Phi_k(r,\theta)|_{\theta=\theta_0}=0 \ , \
i=1,3 \ , \  k=1,3,4 \ .
\end{equation}
\end{subequations}
\section{Flat space solutions}
Flat space is given when $\alpha=0$ and thus  $f=m=l=1$.
\subsection{BPS solutions}
In the Prasad-Sommerfield limit of vanishing potential, i.e. $\beta_2=\beta_3=0$,
explicit BPS solutions in SU(5) have been constructed \cite{meckes}
for $n=1$. They are given by:
\begin{subequations}
\begin{equation}
\Phi_1(r)=\frac{1}{er}\left(\frac{Cr}{\tanh(Cr)}-1\right)
 \ , \ \Phi_3(r)=\frac{2}{\sqrt{3}}\frac{C}{e} \ , \ \ 
\Phi_4(r)=\frac{1}{\sqrt{15}}\frac{C}{e}
\label{bps1}
\end{equation}
\begin{equation}
H_2(r)=H_4(r)=\frac{Cr}{\sinh(Cr)}
\label{bps2}
\end{equation}
\end{subequations}

and all other matter functions identically zero. For $C=1$, the ADM mass of this solution
is given by $m^{C=1}_{ADM}=4\pi\frac{\eta}{e}$. For $C\neq 1$, the mass scales
like $m_{ADM}=Cm^{C=1}_{ADM}=4\pi C\frac{\eta}{e}$. To obtain the solution with the right bcs
for our model, we have to choose $C=\frac{1}{2}\sqrt{\frac{5}{3}}$.
For $n > 1$ and/or $\beta_2\neq 0$, $\beta_3\neq 0$, the solutions have to be
constructed numerically. 
In all our numerical studies however, 
the solution (\ref{bps1}), (\ref{bps2}) proved to be a good starting solution.
\subsection{Numerical results}
We were mainly interested in the effect of the potential. 
We thus first studied the behaviour of the Higgs field function $\Phi_1$ 
for different values of the coupling constants $\beta_2$ and $\beta_3$.
Our results  for the $n=1$ monopole are shown in Fig.~1. For $\beta_2=\beta_3=0$,
$\Phi_1$ is given (analytically) by (\ref{bps1}). For $\beta_2=5$, $\beta_3=0$,
the function still seems to decay power-law like rather than  exponentially, while
for both $\beta_2=0$, $\beta_3=5$ and $\beta_2=\beta_3=5$ the exponential decay is
apparent. From (\ref{mass}), we see that for $\beta_3=0$ only
one of the Higgs fields is massive, while all others remain massless.
This can also be seen by an asymptotic analysis for the case $\beta_3=0$. 
We set 
\begin{equation}
\Phi_{\gamma}=C_{\gamma}+\eta_{\gamma} \ , \ \gamma=1, 3, 4 \ \ \ 
{\rm with} \ \ \
C_{\gamma}=(\frac{\sqrt{5/3}}{2} , \frac{\sqrt{5}}{3} ,\frac{1}{6}) \ .
\end{equation}
The linearised equations for the three Higgs field functions $\Phi_1$, $\Phi_3$
and $\Phi_4$ read (with the prime denoting the derivative with respect to $x$):
\begin{equation}
(x^2 \eta_{\gamma}^{'})^{'}  =  x^2 M_{\gamma\delta}\eta_{\delta} \ \ \
{\rm with} \ \ \ M_{\gamma\delta}=2 C_{\gamma}C_{\delta} \ . 
\end{equation}
The matrix $M_{\gamma\delta}$ has one eigenvector, $C_{\gamma}$ itself
with eigenvalue $2$ and two eigenvectors $C_{\delta_1}$,
$C_{\delta_2}$, the ones orthogonal to $C_{\gamma}$,
with eigenvalues $0$. The solution of the linearised equation is then given by:
\begin{equation}
\eta_{\gamma}=C_{\gamma} e^{-\sqrt{2}x}+ c_1 C_{\delta_1}/x+
c_2 C_{\delta_2}/x \ , \ \ c_1, c_2 \ \ {\rm constant}
\end{equation}
Clearly at large $x$, the power-law decay dominates the behaviour of the
Higgs field functions.

In Fig.~2 we present the mass per winding number of the $n=1$ monopole
and the $n=2$ multimonopole in units of 
$4\pi\frac{\eta}{e}C$, $C=\frac{1}{2}\sqrt{\frac{5}{3}}$
for three different cases: a) $\beta_2\neq 0$, $\beta_3=0$, b) $\beta_3\neq 
0$, $\beta_2=0$ and
c) $\beta_2=\beta_3\neq 0$. Apparently, for all cases, the monopoles are 
in a repulsive phase,
as expected. However, it seems that the mass depends only slightly on 
$\beta_2$. In the case of $\beta_3=0$, the potential can be written as a 
perfect square and thus resembles the potential of the SU(2) case.
However, it should be noted that the present model is not an embedding in 
the sense that the vacuum expectation value has only two non-vanishing 
diagonal entries. We find the following values for the energy per winding number
$E/n$ for the $n=1$ monopole and the $n=2$ multimonopole for $\beta_2$ (respectively
$\beta_3$) $\rightarrow\infty$:

\begin{center}
Table 1 \\
\medskip
\begin{tabular}{|l|l|l|}
\hline 
$ $ & $n=1$ & $n=2$ \\
\hline \hline
$\beta_2\rightarrow \infty$, $\beta_3=0$ & $1.076$ & $1.081$ \\
\hline
$\beta_3\rightarrow \infty$, $\beta_2=0$ & $1.650$ & $2.112$ \\
\hline
$\beta_2=\beta_3\rightarrow\infty$ & $1.657$ & $2.135$ \\
\hline
\end{tabular}
\end{center}
These can be compared to the ones computed in the SU(2) case, where
the mass per winding number has been obtained for $\tilde{\lambda} \rightarrow \infty$, where
$\tilde{\lambda}$ is the Higgs self-coupling constant of the standard 
SU(2) Higgs potential $\tilde{V}=\frac{\tilde{\lambda}}{4}
(\tilde{\Phi}^2-\tilde{\eta})^2$ \footnote{In the following, we denote
all fields, constants etc. of the SU(2) case with a tilde.}. It was found that $E/n(\tilde{\lambda}\rightarrow\infty)$
is equal to $1.787$ for $n=1$ and equal to $2.293$ for $n=2$ \cite{marinov,kkt}. Comparing this,
we see that the order of magnitude agrees with our results. 

\section{Gravitating solutions}
\subsection{Spherically symmetric solutions}
In the case of spherical symmetry ($n=1$), the gauge field functions $H_1$, $H_3$
and the Higgs field function $\Phi_2$ are identically zero. In addition 
$m=l$, $H_2=H_4$
and all functions depend only on $x$. We here give the explicit form
of the equations to demonstrate the changes in comparison to the
SU(2) case. For the axially symmetric case, these changes are analog. 
The equations for the remaining functions read (renaming $H_2=H_4\equiv 
K$ and the prime
denoting the derivative with respect to $x$):
\begin{equation}
\frac{f^{''}}{f}+2\frac{f^{'}}{f x}+\frac{1}{2}\frac{f^{'}l^{'}}{f l}-
\frac{(f^{'})^2}{f^2}=2\alpha^2 \left(\frac{l}{f}U(\Phi_1,\Phi_3,\Phi_4)
+\frac{f}{lx^4}(K^2-1)^2+2\frac{f}{lx^2}(K^{'})^2    \right)
\end{equation}
\begin{equation}
\frac{1}{2}\frac{l^{''}}{l}+\frac{3}{2}\frac{l^{'}}{l x}-
\frac{1}{4}\frac{(l^{'})^2}{l^2}=2\alpha^2 \left(\frac{l}{f}U(\Phi_1,\Phi_3,\Phi_4)+
\frac{f}{lx^2}(K^{'})^2-\frac{K^2 \Phi_1^2}{x^2}    \right)
\end{equation}
with
\begin{eqnarray}
& & U(\Phi_1,\Phi_3,\Phi_4)=-\frac{\beta_2^{2}}{2}\left(\Phi_1^2+\Phi_3^2+\Phi_4^2-1\right)^2\nonumber \\
&-&
\beta_3^2 \left(
\Phi_1^2(\frac{\Phi_1^2}{4}+\frac{9}{10}\Phi_4^2-\frac{7}{30})
 + \Phi_3^2(\frac{\Phi_3^2}{4}+\frac{2}{5}\Phi_4^2-\frac{7}{30}
-\frac{4\sqrt{5}}{30}\Phi_3\Phi_4) +\frac{7}{60} (\Phi_4^2-1)^2\right)
\end{eqnarray}
for the metric functions $f$ and $l$ and
\begin{equation}
\left(\frac{f}{\sqrt{l}} K^{'} \right)^{'}=\frac{f}{\sqrt{l}}K
\left(\frac{K^2-1}{x^2}+ \Phi_{1}^{2} \frac{l}{f} \right)
\end{equation}
\begin{equation}
\left(x^2 \sqrt{l} \Phi_1^{'}\right)^{'}=
\frac{l^{3/2} x^2 \Phi_1}{f}\left[\beta_2^2 \left(\Phi_1^2 + \Phi_3^2  +
 \Phi_4^2-1 \right) 
+ \beta_3^2 \left(\frac{\Phi_1^2}{2}  + \frac{27}{30} \Phi_4^2 - 
\frac{7}{30} \right) \right] + 2 K^2 \sqrt{l}\Phi_1
\end{equation}
\begin{equation}
\left(x^2 \sqrt{l} \Phi_3^{'}\right)^{'}=
\frac{l^{3/2} x^2 \Phi_3}{f}\left[\beta_2^2 \left(\Phi_1^2 + \Phi_3^2  +
 \Phi_4^2-1 \right) 
+ \beta_3^2 \left(\frac{\Phi_3^2}{2}  + \frac{2}{5} \Phi_4^2 - 
\frac{1}{\sqrt{5}} \Phi_3 \Phi_4 -\frac{7}{30} \right) \right]
\end{equation}
\begin{eqnarray}
\left(x^2 \sqrt{l} \Phi_4^{'}\right)^{'}&=&
\frac{l^{3/2} x^2 \Phi_4}{f}\left[\beta_2^2 \left(\Phi_1^2 + \Phi_3^2  +
 \Phi_4^2-1 \right) \right.  \nonumber \\ 
 &+& \left.  \beta_3^2 \left(-\frac{\Phi_3^3}{3\sqrt{5}\Phi_4}  +
 \frac{9}{10} \Phi_1^2 +
\frac{2}{5} \Phi_3^2 +\frac{7}{30}(\Phi_4^3-1) \right) \right]
\end{eqnarray}
for the gauge field function $K$ and the three Higgs field functions $\Phi_i$, $i=1,3,4$.
In the case of gravitating solutions, i.e. $\alpha > 0$, the solutions
in the limit of $\beta_2=\beta_3=0$ have constant functions 
$\Phi_3$ and $\Phi_4$. We thus obtain for the mass
exactly that obtained for the SU(2) monopoles \cite{gsu2} when we express
it in units of $\frac{1}{2}\sqrt{\frac{5}{3}}$.
Note, though, that in order to be able to compare the results obtained, 
we have to rescale $\alpha\rightarrow \frac{\alpha}{2}\sqrt{\frac{5}{3}}$.

In order to demonstrate the influence of the Higgs coupling constants
$\beta_2$, $\beta_3$, we show in Fig.s 3a and 3b the value
of the Higgs field function $\Phi_3$ and $\Phi_4$, respectively, at the origin 
in dependence on the parameters $\beta_2$, $\beta_3$ for $\alpha=1$.
For comparison, the corresponding values in flat space ($\alpha=0$) 
are also shown. Since in the BPS limit ($\beta_2=\beta_3=0$), the functions
$\Phi_3$ and $\Phi_4$ are constant, all curves meet at $\frac{\sqrt{5}}{3}$ for
$\Phi_3$ and $\frac{1}{6}$ for $\Phi_4$, respectively, in this limit.
Again, it is apparent from these two figures that the parameter
$\beta_2$ does not have a large influence. The functions $\Phi_3$ and $\Phi_4$
become monotonically decreasing functions of $x$ for the 
choices of $\beta_2$, $\beta_3$ shown in Fig.s 3a, 3b starting at the value shown
with zero derivative and decreasing to their asymptotic values. The inclusion
of gravity leads to an increase of $\Phi_4$ at $x=0$, while
the value $\Phi_3(x=0)$ seems to depend only slightly
on the parameter $\alpha$.

In order to be able to study further details of
the $n=1$ solutions and be able to compare them with the SU(2) case,
we have adopted the Schwarzschild-like ansatz for the metric used
in \cite{gsu2,lw} rather than the isotropic ansatz (\ref{metric}):
\begin{equation}
ds^2=-A^2(r_s)N(r_s)dt^2+N^{-1}(r_s)dr_s^2+r_s^2(d\theta^2+\sin^2\theta d\varphi^2) \ .
\end{equation}
The isotropic coordinate $r$ is relate to the Schwarzschild-coordinate $r_{s}$ through
the following transformation \cite{hkk2}:
\begin{equation}
\frac{dr}{dr_s}=\frac{1}{\sqrt{N(r_s)}}\frac{r}{r_s}
\end{equation}
and thus can - for generic cases - only be obtained numerically.

In the SU(2) case, it was observed that the gravitating monopole solutions
exist only up to a maximal value of the gravitational coupling 
$\tilde{\alpha}$, $\tilde{\alpha}_{max}$ \cite{gsu2}. $\tilde{\alpha}_{max}$ 
is a decreasing
function of the Higgs self-coupling constant starting from 
$\tilde{\alpha}_{max}=1.403$ in the BPS limit. 
For bigger $\tilde{\alpha}$, the Schwarzschild radius becomes larger than the
radius of the monopole core. Consequently,
an extremal horizon $\tilde{x}_{h_s}$ forms with $N(\tilde{x}_{h_s})=
\partial_{\tilde{x}} N|_{\tilde{x}=
\tilde{x}_{h_s}}=0$ and in the limit $\tilde{\alpha}\rightarrow
\tilde{\alpha}_{max}$ the solutions bifurcate with the branch of extremal
Reissner-Nordstr\"om solutions. The solutions outside the horizon $\tilde{x}_{h_s}$
are described by this solutions, while in the interval $0\leq \tilde{x}\leq 
\tilde{x}_{h_s}$ they are
non-trivial and non-singular.
For small Higgs self-coupling a second branch
of solutions was found extending backwards
from $\tilde{\alpha}_{max}$ to $\tilde{\alpha}_{cr} < \tilde{\alpha}_{max}$ with
$\tilde{\alpha}_{cr}=1.386$ in the BPS limit. The solutions then reach
their limiting solution for $\tilde{\alpha}\rightarrow
\tilde{\alpha}_{cr}$.
For intermediate Higgs masses, a new phenomenon
was observed \cite{lw}. A second, inner minimum
appears that drops down much quicker to zero
than the outer one. The limiting solution
thus represents a non-abelian black hole. 
In the table below, we give $\alpha_{max}$,
$\alpha_{cr}$ for $\beta_2=0$ in units of $\frac{1}{2}\sqrt{\frac{5}{3}}$ in order
to be able to compare it with the SU(2) results. We also give the value of the outer minimum $N^o_{min}$ at the appearance of
the second, inner minimum, if this phenomenon is observed. 
\begin{center}
Table 2 \\
\medskip
\begin{tabular}{|l|l|l|l|}
\hline 
$\beta_3 $ & $\frac{1}{2}\sqrt{\frac{5}{3}}\cdot\alpha_{max}$ & $\frac{1}{2}\sqrt{\frac{5}{3}}\cdot\alpha_{cr}$
& $N^o_{min}$  \\
\hline \hline
$0$ & $1.403$ & $1.386$ &$ -$\\
\hline
$0.2$ & $1.370$ & $1.366$ &$ -$\\
\hline
$0.3$ & $1.350$ & $1.348$ &$ -$\\
\hline
$0.5$ & $1.308$ & $1.308$ &$ -$\\
\hline
$1$ & $1.223$ & $1.223$& $ -$ \\
\hline
$2$ & $1.117$ & $1.117$ &$ -$\\
\hline
$3$ & $1.050$ & $1.050$ &$ -$\\
\hline
$4$ & $1.007$ & $1.007$ &$ -$\\
\hline
$5$ & $0.974$ & $0.974$ &$ -$\\
\hline
$7$ & $0.926$ & $0.926$ &$ -$\\
\hline
$10$ & $0.884$ & $0.884$ &$ -$\\
\hline
$15$ & $0.534$ & $0.534 $ &$ 0.24\cdot 10^{-9}$ \\
\hline
$16$ & $0.851$ & $0.851$ &$ 0.49\cdot 10^{-7}$\\
\hline
$17$ & $0.845$ & $0.845$ &$ 0.94\cdot 10^{-6}$\\
\hline
$20$ & $0.829$ & $0.829$ &$ 0.37\cdot 10^{-4}$\\
\hline
$50$ & $0.769$ & $0.769$ &$0.80\cdot 10^{-2} $\\
\hline
$100$ & $0.746$ & $0.746$ &$ 0.21\cdot 10^{-1}$\\
\hline
\end{tabular}
\end{center}
Clearly, the phenomenon observed for
intermediate Higgs masses in the SU(2) model \cite{lw} 
is observed here for $\beta_3 \geq 15$, while a second branch of solutions exists 
only for $\beta_3 \leq 0.3$. For all other $\beta_3$ we find that
$\alpha_{max}=\alpha_{cr}$.

Analogous to the SU(2) case, the gravitating monopoles bifurcate
with the branch of extremal Reissner-Nordstr\"om (RN) solutions in the limit
$\alpha\rightarrow\alpha_{max}$ (resp. $\alpha\rightarrow\alpha_{cr}$).
The approach to criticality is shown for the two Higgs fields  
$\Phi_3$, $\Phi_4$ in Fig.s 4a, 4b. for $\beta_2=0$, $\beta_3=0.3$.
$\Phi_3$ and $\Phi_4$ are shown as functions of the dimensionless 
Schwarzschild-coordinate $x_s=e\eta r_s$ for four different values
of the gravitational coupling $\alpha$. Clearly, the limiting solution
is not yet reached for $\alpha_{max}$, but rather for 
$\alpha_{cr} < \alpha_{max}$, the numerical values of which are given 
in Table 2. An extremal horizon $x_{h_s}$ forms and $\Phi_3$
as well as $\Phi_4$ are constant and equal to their asymptotic
values for $x_s \geq x_{h_s}$, while they are non-trivial for 
$x$ $\epsilon$ $[0:x_{h_{s}}]$. 
This is very similar
to what was observed previously in the SU(2) case for the remaining
matter field functions. Striking is that the two functions
look qualitatively very similar in their approach to criticality.

Looking at the results of flat space, we expect that a similar pattern exists for the case
of $\beta_2=\beta_3\neq 0$, while for $\beta_2\neq 0$, $\beta_3=0$, we find that
$\alpha_{max}$ depends only very little on $\beta_2$.

\subsection{Axially symmetric solutions}
While in flat space SU(2) monopoles as well as SU(5) monopoles - as was demonstrated
by us in this paper - are either repelling or non-interacting
(in the BPS limit), it was shown in the SU(2) case \cite{hkk} that
gravity is able to overcome the repulsion for sufficiently small Higgs boson
masses. We thus limit our analysis of axially symmetric SU(5) monopoles
to this important point. We find that for a fixed $\alpha$ the values
of $\beta^{eq}_2$, $\beta^{eq}_3$ for which the mass of the $n=1$ monopole
is equal to the mass per winding number of the $n=2$ multimonopole
roughly forms a half-circle in the $\beta_2$-$\beta_3$-plane. E.g. we find
for $\alpha=1.4$, that for $\beta_3=0$ the value of $\beta^{eq}_2\approx 0.045$,
while for $\beta_2=0$ we have $\beta^{eq}_3\approx 0.043$ and for
$\beta^{eq}_2=\beta^{eq}_3\approx 0.03$. This suggests that
\begin{equation}
\sqrt{\beta^{eq}_2(\alpha)+\beta^{eq}_3(\alpha)}\approx C_1(\alpha) \ \ , \ \
{\rm with}\ \  C_1 \ \ {\rm constant} \ .
\end{equation}
We find that $C_1$ is an increasing function of $\alpha$, e.g. we find that
$C_1(\alpha=1.6)\approx 0.056$ and $C_1(\alpha=1.8)\approx 0.2$. This latter
value of $C_1$ for $\alpha=1.8$ suggests in comparison with the other two that
$C_1$ is in fact a strongly increasing function of $\alpha$. This is indeed
what was also found in the SU(2) case \cite{hkk}. Moreover, the order of 
magnitude of $\beta^{eq}_2$,  $\beta^{eq}_3$ agrees with what was found in the
SU(2) case. This thus suggests that also in the SU(5) case, bound multimonopoles
only exist for sufficiently small values of the Higgs self-coupling constants. 

\section{Non-abelian black holes}
For the numerical construction of non-abelian black holes, we followed \cite{hkk2} and
introduced new metric functions $f_2$, $l_2$ and $m_2$ 
\begin{equation}
f_2=\frac{f}{z^2} \ , \ \ l_2=\frac{l}{z^2} \ , \ \ m_2=\frac{m}{z^2} \ ,
\end{equation}
where $z=1-\frac{x_h}{x}$ is the compactified coordinate that
maps the infinite interval $[x_h:\infty[$ to the finite interval
$[0:1]$.  
The new boundary conditions at the horizon $x_h$ now read \cite{hkk2}:
\begin{equation}
\left(f_2-\frac{\partial f_2}{\partial z}\right)|_{z=0}=0 \ , \ \ 
\left(m_2+\frac{\partial m_2}{\partial z}\right)|_{z=0}=0 \ , \ \ 
\left(l_2+\frac{\partial l_2}{\partial z}\right)|_{z=0}=0 \ .
\end{equation}
The isotropic horizon $x_h$ is not a physical quantity. We thus determine
for every solution obtained the parameter $x_{\Delta}$ given by:
\begin{equation}
x_{\Delta}=\sqrt{\frac{A}{4\pi}} \ \ {\rm with} \ \ A= 2\pi \int_{0}^{\pi} d\theta
\sin\theta x_h^2  \frac{\sqrt{l_2(x=x_h,\theta)
m_2(x=x_h,\theta)}}{f_2(x=x_h,\theta)} \ .
\end{equation}
$A$ is the area of the horizon and directly related to the entropy $S$ of the
black hole: $S=\frac{A}{4}$. Note that the Schwarzschild-like horizon $x_{h_{s}}$ 
is exactly defind to be the radius of the horizon with area $A$.

\subsection{Spherically symmetric solutions}
Spherically symmetric black hole solutions can be constructed for
$n=1$. Similarly as in the case of regular solutions they have $m_2=l_2$,
$H_2=H_4$, $H_1$=$H_3$=$\Phi_2=0$ and all remaining functions depend only on $x$.
In fact the non-abelian black hole solutions can equally well be constructed 
using the Ansatz of the Schwarschild-like metric. 
The boundary conditions are then directly
imposed at $x = x_{h_{s}}= x_{\Delta}$.

We find that the domain of existence of the spherically
symmetric non-abelian black holes in the $\alpha$-$x_{\Delta}$-plane 
is of similar shape than the domain of existence in the SU(2) \cite{gsu2}
and in the SU(3) \cite{bp} case, respectively. Fixing $\alpha$ and increasing
$x_{\Delta}$, the non-abelian solutions exist for
$x_{\Delta} \leq x_{\Delta}^{max}$ and at $ x_{\Delta} = x_{\Delta}^{cr}$  bifurcate 
with the branch of 
non-extremal Reissner-Nordstr\"om (RN) solution for small $\alpha$ and 
with the branch of extremal Reissner-Nordstr\"om (RN) solution for large $\alpha$, respectively.
In this paper, we put emphasis on the
case $\alpha\cdot C = 0.5$, $C=\frac{1}{2}\sqrt{\frac{5}{3}}$,
for which  the solutions bifurcate with the branch of
non-extremal RN solutions.
For $\beta_2 = \beta_3 = 0$ the bifurcation
occurs on a second branch of solutions
which exists for  $x_{\Delta}^{cr}\leq x_{\Delta} \leq x_{\Delta}^{max}$, i.e.
that in this interval two  different
black hole solutions can be constructed. For $\beta_2 = \beta_3$ sufficiently large
only one branch of solutions exists and $x_{\Delta}^{max}=x_{\Delta}^{cr}$. 
This is demonstrated in Fig. 5 for $n=1$ and $\beta_2 = \beta_3 = 0$, resp. $\beta_2 = \beta_3=1$.
We show the energy $E(1)$ of the non-abelian solutions and that of the corresponding RN solutions.
The occurence of a second branch of solutions
for $\beta_2 = \beta_3 = 0$ is apparent from this figure with
$x_{\Delta}^{max}\approx 1.064$ and $x_{\Delta}^{cr}\approx 0.936$. For
$\beta_2 = \beta_3=1$ only one branch of solutions exists and we find that
$x_{\Delta}^{max}=x_{\Delta}^{cr}\approx 0.9676$.

A good indication for the bifurcation with the branch of non-extremal RN solutions
is the fact that in the limit
$x_{\Delta}\rightarrow x_{\Delta}^{cr}$ 
the values of the matter fields at the horizon tend  to their asymptotic values.
Since these functions
are monotonic functions of $x$, they obviously 
become constant on the whole interval $[x_{\Delta}:\infty[$ in the critical limit.
We demonstrate this in Fig. 6 for $H_2(x_{\Delta})$ and
$\Phi_a(x_{\Delta})$, $a=1,3,4$.
The value of the gauge function at the horizon, $H_2(x_{\Delta})$, tends to zero,
while the values of the three Higgs field function at the horizon, $\Phi_a(x_{\Delta})$,
approach their respective asymptotic values (see (\ref{higgs})). 
    
\subsection{Axially symmetric solutions}
In the SU(2) case, it was found \cite{hkk2} that the critical values of the horizon parameter $x_{\Delta}$
increase with increasing $n$. However, it was also shown that the qualitative shape of
the domain of existence doesn't change. We observe the same for the SU(5) case.
This is demonstrated in Fig. 5, where we shown together with the energy of the $n=1$ solutions
the energy $E(2)$ of the axially symmetric black hole solutions for
two different values of $\beta_2=\beta_3$. Clearly, the solutions exist for
higher values of $x_{\Delta}$. We find for $\beta_2=\beta_3=0$ that 
$x_{\Delta}^{max}\approx 2.158$ and $x_{\Delta}^{cr}\approx 1.667$, while for $\beta_2=\beta_3=1$, we have
$x_{\Delta}^{max}=x_{\Delta}^{cr}\approx 1.691$. Indeed, this figure demonstrates again the occurence
of a second branch of solutions for $\beta_2=\beta_3=0$, while for $\beta_2=\beta_3=1$ only one branch of
non-abelian black hole solutions exists.

As an indication of the deformation of the horizon, which is defined to be the
surface of constant $x_h$,  we have computed the circumference
of the horizon along the equator $L_e$
\begin{equation}
L_e=\int_{0}^{2\pi} d\varphi \sqrt{g_{\varphi\varphi}} \mid_{x=x_h, \theta=\pi/2}=
2\pi x_h \sqrt{\frac{l(x=x_h, \theta=\pi/2)}{f(x=x_h, \theta=\pi/2)}}
\end{equation}
and the circumference of the horizon along the poles $L_p$:
\begin{equation}
L_p=2\int_{0}^{\pi} d\theta \sqrt{g_{\theta\theta}}\mid_{x=x_h, \varphi=const.}
=2 x_h \int_{0}^{\pi} d\theta \sqrt{\frac{m(x=x_h, \theta)}{f(x=x_h, \theta)}} \ .
\end{equation}
It is apparent that the ratio $L_e/L_p$ is equal to one
for the spherically symmetric solutions since $l=m$ for $n=1$ and the angle dependence disappears.
For the axially symmetric solutions, however, the ratio deviades from one. This is demonstrated
in Fig. 7, where we show $L_e/L_p$ as function of $x_{\Delta}$ for the
$n=2$ non-abelian black hole with $\alpha\cdot C=0.5$, $C=\frac{1}{2}\sqrt{\frac{5}{3}}$.
The case $\beta_2=\beta_3$ is the analog of what was studied in the SU(2) case for $\tilde{\lambda}=0$.
The ratio reaches a minimum at a value of $x_{\Delta}$ close to the maximal $x_{\Delta}$, then on the second branch
of solutions reaches a maximal value which is bigger than one and then drops down to one
at $x_{\Delta}=x^{cr}_{\Delta}$. For non-vanishing Higgs coupling constants (we have chosen $\beta_2=\beta_3=1$),
the apparent effect is the decrease of the minimal and the increase of the maximal value of $L_e/L_p$. 
Thus the inclusion of the Higgs potential increases the deformation of the regular horizon 
of the non-abelian black holes. The qualitative feature changes in the sense that now no second branch of
solutions exists and thus the local extrema of the curve are reached for $x_{\Delta} < x^{cr}_{\Delta}$.
Moreover, we observe a sharp drop of the ratio from its maximal value to the value one 
at $x_{\Delta} = x^{cr}_{\Delta}$. Fig. 7 leads us further to the conclusion that
the observation of two local extrema (a minimum and a maximum) for the ratio $L_e/L_p$ 
seems a generic feature for all Higgs coupling constants as long as the critical solutions
into which the non-abelian black holes merge at $x_{\Delta} = x^{cr}_{\Delta}$ 
are non-extremal Reissner-Nordstr\"om solutions.

\section{Summary and Conclusions}
Since it is believed that monopoles have formed in the early universe through the spontaneous
symmetry breakdown of a Grand Unified Theory (GUT) down to a gauge group containing 
a $U(1)$ subgroup, the study
of monopoles arising in SU(5) Yang-Mills-Higgs theory which (through an appropriate potential)
is spontaneously broken down
to $SU(3)\times SU(2) \times U(1)$ seems of importance.

We have thus studied spherically as well as axially symmetric solutions in
SU(5) Yang-Mills-Higgs theory both in flat and curved space. 
Concerning the globally
regular solutions, we find that in flat space 
very similar to the corresponding solutions in SU(2) Yang-Mills-Higgs theory,
the monopoles are either non-interacting or repelling. Moreover, we find
that the order of magnitude of the mass in the limit
of  the Higgs self-coupling constants going to infinity agrees 
with the SU(2) results. Minimal coupling of gravity to the system leads 
in principle to two different types of non-abelian solutions: 
a) globally regular, gravitating
monopoles and b) non-abelian black holes. We find that
the qualitative features like the appearance of a second branch of solutions
for small Higgs self-coupling constants or the appearance of a new phenomenon
for intermediate values of the Higgs self-coupling constants are also found in
this system. Since these phenomena have also been observed for spherically
symmetric solutions in SU(3) Einstein-Yang-Mills-Higgs theory (EYMH) \cite{bp}, it is likely
that what was observed in SU(2) seems to be a generic feature
of SU(N) EYMH theory with the Higgs field in the adjoint representation.

We also studied spherically and axially symmetric non-abelian black holes
solutions.
The corresponding solutions for the SU(2) gauge group were studied in great
detail in \cite{hkk2} for vanishing Higgs coupling constant (BPS limit).
It seems, however, that the ``physically" relevant solutions related to SU(5)
have to be constructed  for non-vanishing values of the Higgs self-coupling
constants $\beta_2$, $\beta_3$. 
That's why we have put 
emphasis on the influence of a non-trivial Higgs potential 
on the solutions.  Of course, the problem is numerically involved
since 
 the equations have to be solved for four continuous parameters:
$\beta_2$, $\beta_3$, $\alpha$ and $x_h$. In addition there is 
a discrete parameter, namely the winding number $n$. Nevertheless, we 
have tried to determine the main features of the influence of the
potential on the solutions. We find that:
(i) the parameters $\beta_{2}$ and $\beta_3$
effect in a non negligible way the domain of existence
in the $\alpha$-$x_{\Delta}$-plane,
(ii) for $\beta_2$, $\beta_3$ large enough, the solutions
bifurcate into a non-extremal Reissner-Nordstr\"om solution
on the first branch of solutions 
without a backbending in the parameter $x_{\Delta}$, 
(iii) the deformation of the horizon
of the axially symmetric black hole solutions increases 
for $\beta_2$, $\beta_3 \neq 0$,  
(iv) the surface gravity - at least for small values of $\beta_2$, $\beta_3$ -
depends only weakly  on $\beta_2$, $\beta_3$.\\
\\
\\
 
{\bf Acknowledgements}
Y. B. is grateful to the Belgian F. N. R. S. for financial support.
B. H. was supported by the EPSRC. We thank CERN for financial support
and its hospitality.

\newpage
\begin{fixy}{-1}
\begin{figure}
\centering
\epsfysize=12cm
\mbox{\epsffile{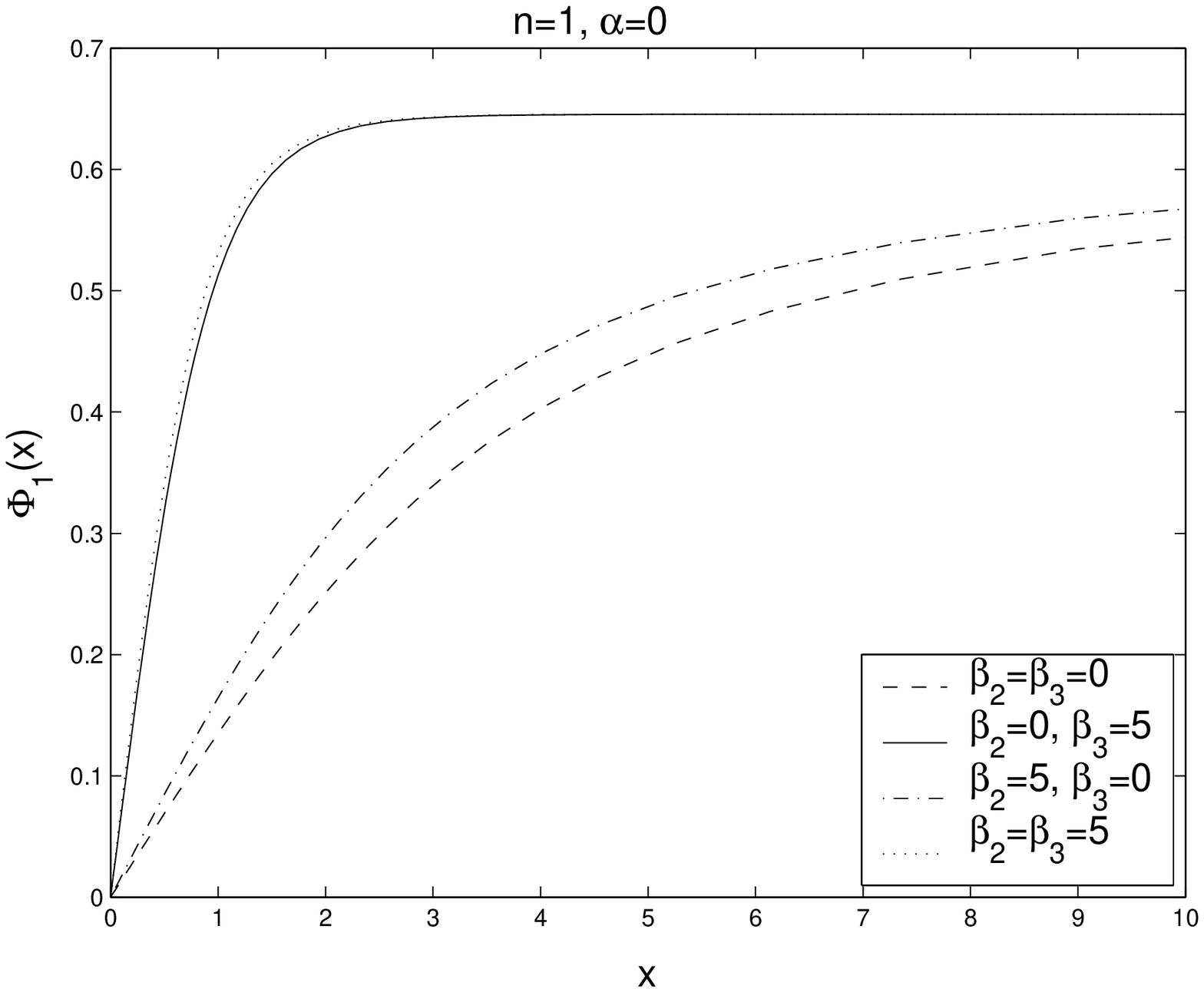}}
\caption{The Higgs field function $\Phi_1(x)$ is shown as function of $x$
for $n=1$, $\alpha=0$ and four different choices of the coupling constants
$\beta_2$, $\beta_3$.
 }
\end{figure}
\end{fixy}
 
\newpage
\begin{fixy}{-1}
\begin{figure}
\centering
\epsfysize=12cm
\mbox{\epsffile{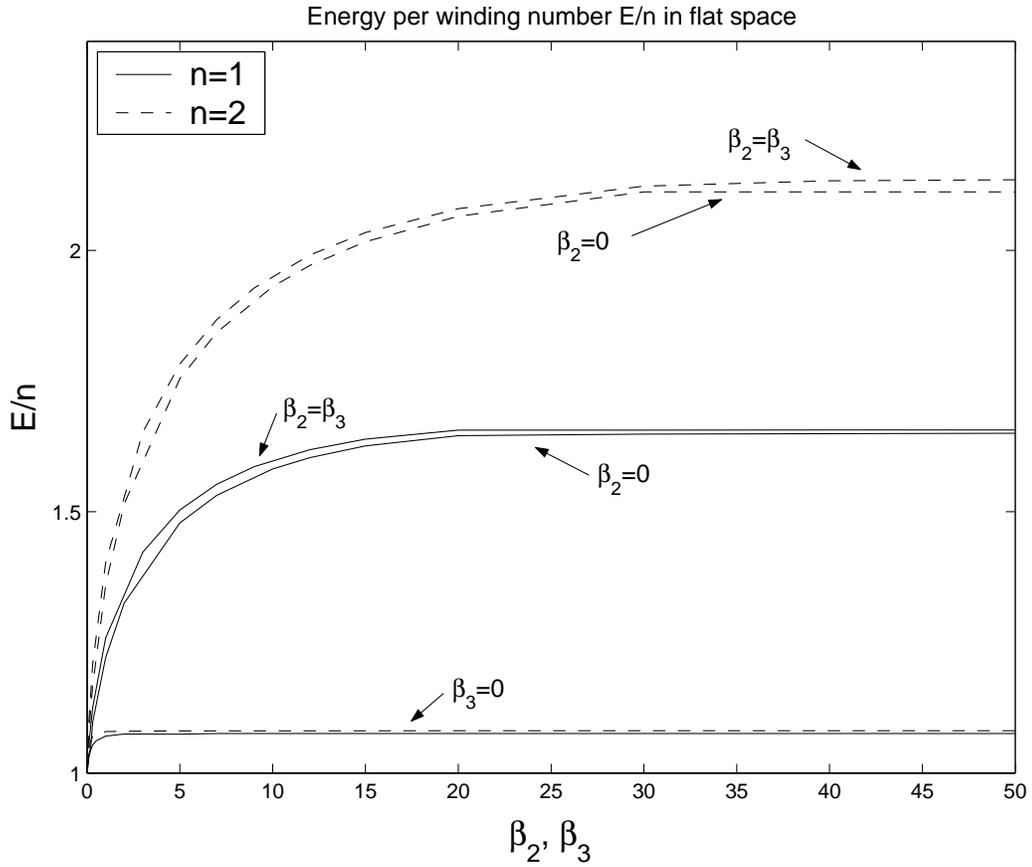}}
\caption{The energy per winding number $E/n$ in
 units of $4\pi\frac{\eta}{e}C$, $C=\frac{1}{2}\sqrt{\frac{5}{3}}$ is shown
for the $n=1$ monopole (solid)  and the $n=2$ multimonopole (dashed)
for $\beta_3=0$, respectively $\beta_2=0$,  respectively $\beta_2=\beta_3$, 
as function of $\beta_2$, respectively $\beta_3$.
 }
\end{figure}
\end{fixy}
\newpage
\begin{fixy}{0}
\begin{figure}
\centering
\epsfysize=10cm
\mbox{\epsffile{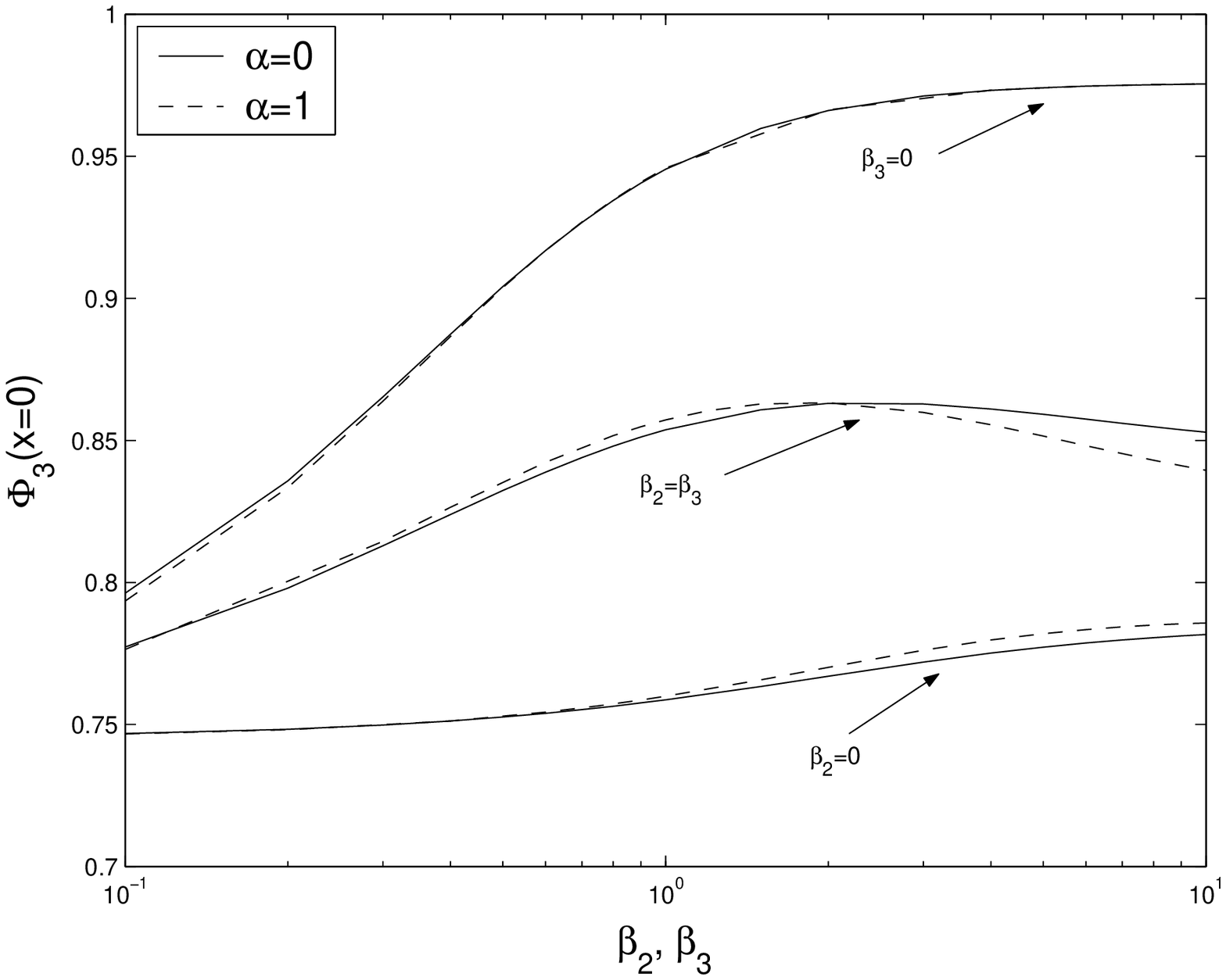}}
\caption{\label{Fig.3a} The value of the Higgs field function
$\Phi_3(x)$ at the origin, 
$\Phi_3(x=0)$, is shown for the $n=1$ monopole
as function of $\beta_2$ ($\beta_3$, respectively)
for $\alpha=0$ and $\alpha=1$, respectively.}
\end{figure}

\begin{figure}
\centering
\epsfysize=10cm
\mbox{\epsffile{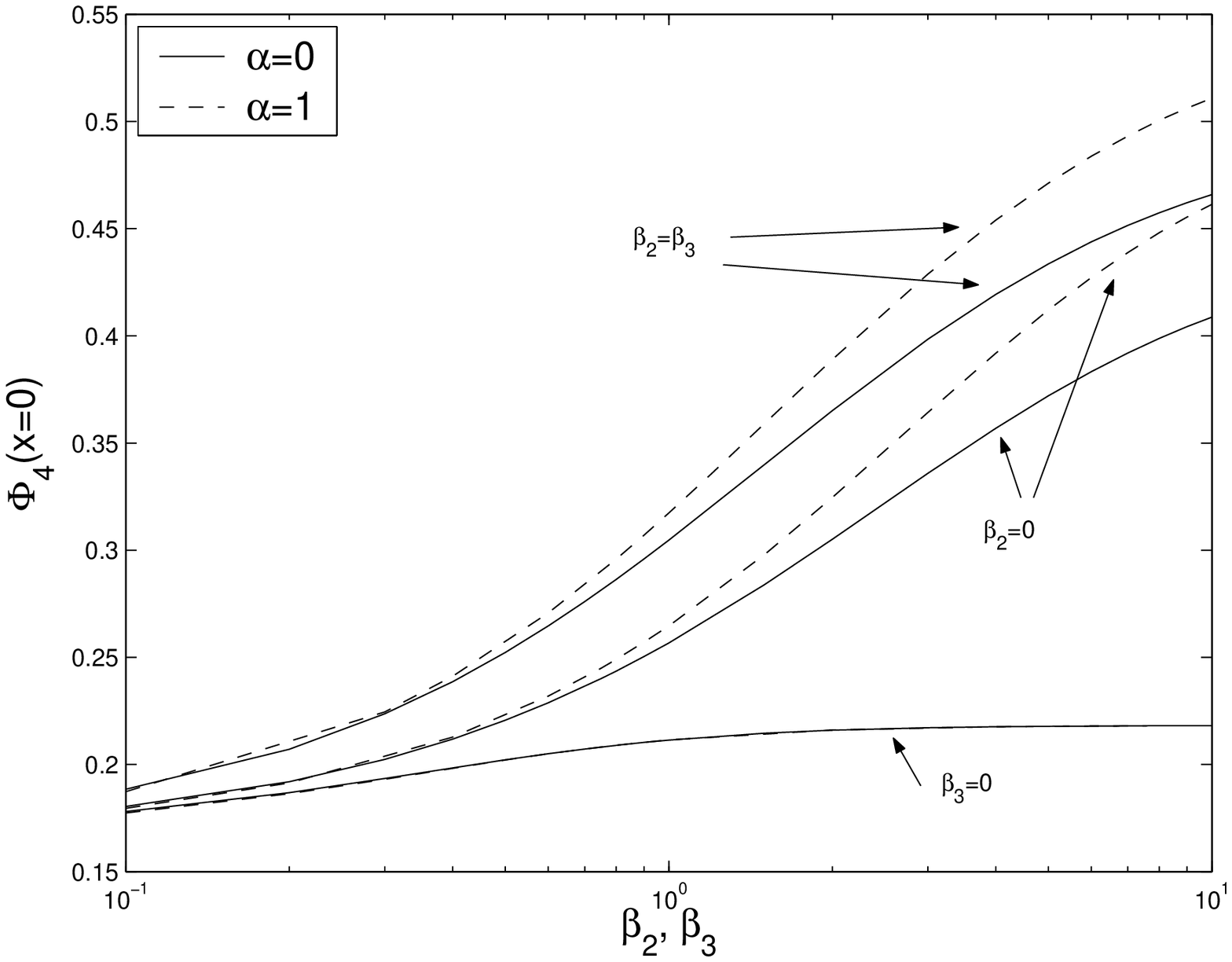}}
\caption{\label{Fig.3b}The value of the Higgs field function
$\Phi_4(x)$ at the origin, 
$\Phi_4(x=0)$, is shown for the $n=1$ monopole
as function of $\beta_2$ ($\beta_3$, respectively)
for $\alpha=0$ and $\alpha=1$, respectively.  }
\end{figure}
\end{fixy}
\newpage

\begin{fixy}{0}
\begin{figure}
\centering
\epsfysize=10cm
\mbox{\epsffile{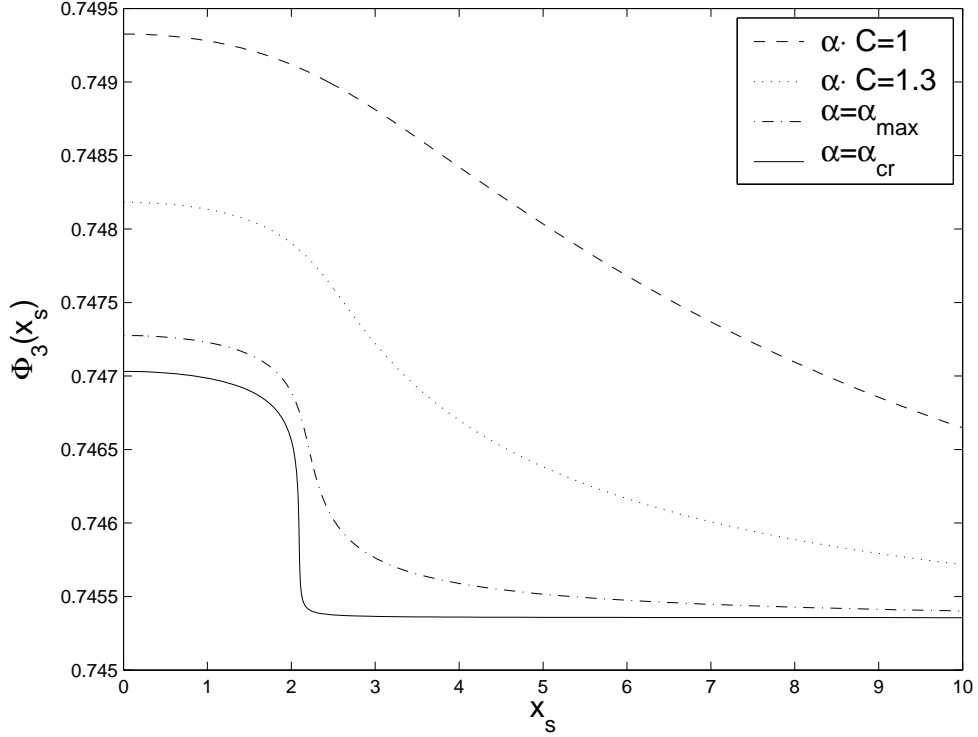}}
\caption{\label{Fig.4a} The  Higgs field function
$\Phi_3(x_s)$  is shown for the $n=1$ monopole
as function of the dimensionless Schwarzschild
coordinate $x_s=e\eta r_s$ for $\beta_2=0$, $\beta_3=0.3$
for four different values of $\alpha$ (with $C=\frac{1}{2}\sqrt{\frac{5}{3}}$)
including $\alpha=\alpha_{cr}$.}
\end{figure}

\begin{figure}
\centering
\epsfysize=10cm
\mbox{\epsffile{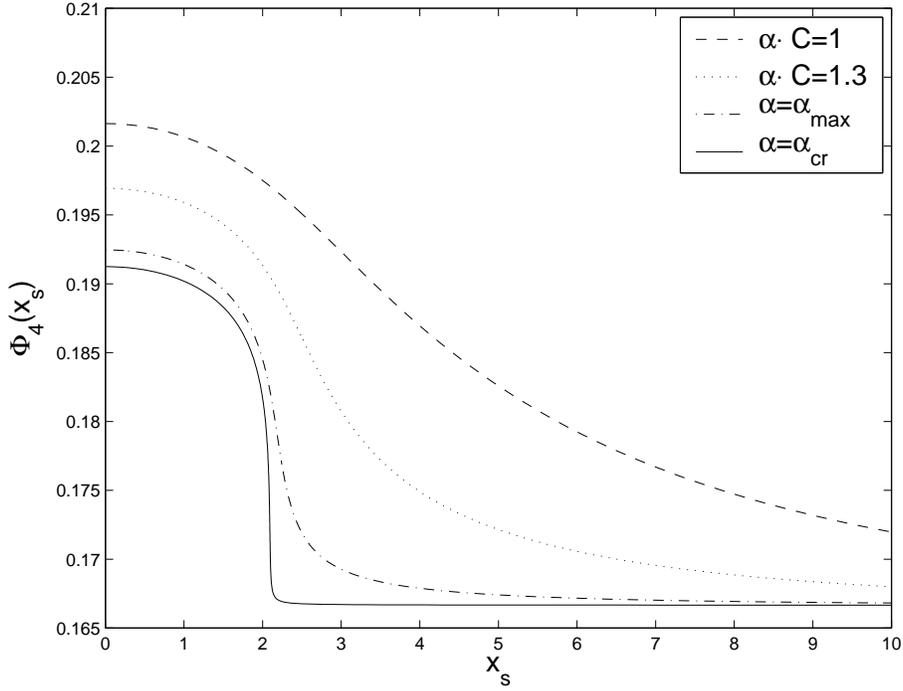}}
\caption{\label{Fig.4b}Same as Fig. 4a but for the Higgs field
function $\Phi_4(x_s)$.  }
\end{figure}
\end{fixy}

\begin{fixy}{-1}
\begin{figure}
\centering
\epsfysize=12cm
\mbox{\epsffile{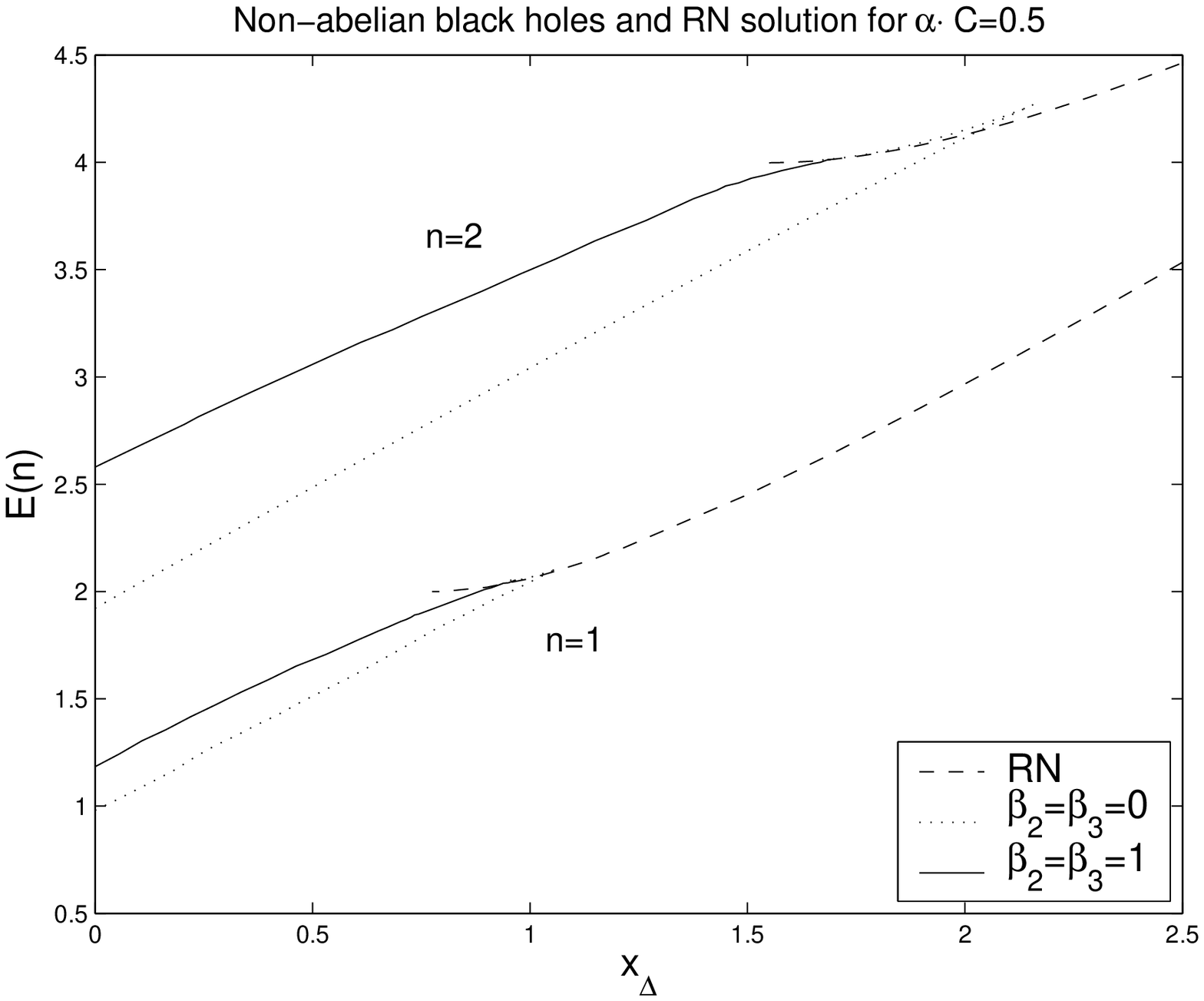}}
\caption{\label{Fig.5}The energy $E(n)$ in units of 
$4\pi\frac{\eta}{e}C$, $C=\frac{1}{2}\sqrt{\frac{5}{3}}$ is shown 
for the $n=1$ and the $n=2$ non-abelian black hole as function of $x_{\Delta}$
for two values of $\beta_2=\beta_3$ and $\alpha\cdot C=0.5$. For comparison,
also the energy of the corresponding Reissner-Nordstr\"om (RN) solution is shown.}
\end{figure}
\end{fixy}

\begin{fixy}{-1}
\begin{figure}
\centering
\epsfysize=12cm
\mbox{\epsffile{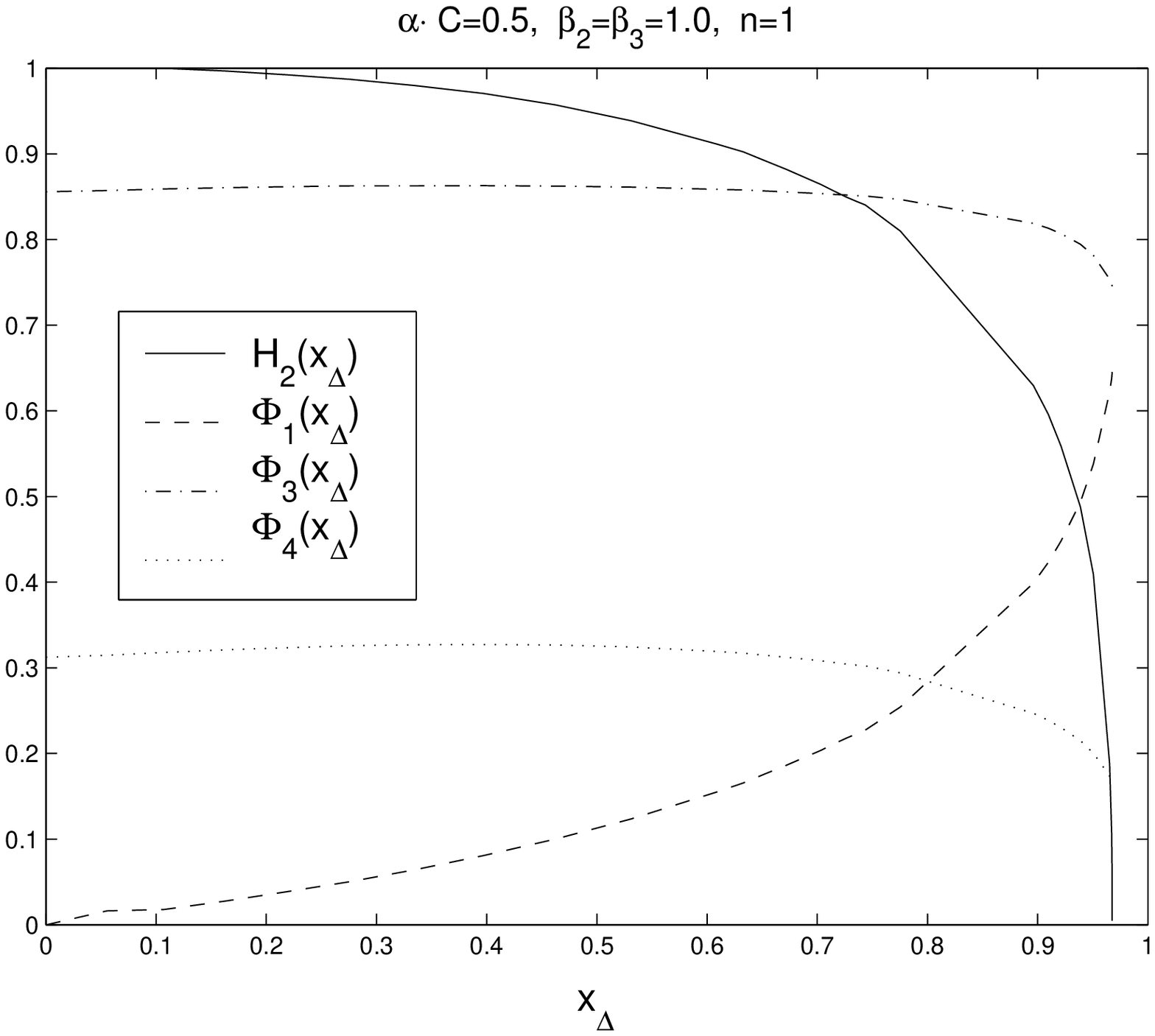}}
\caption{\label{Fig.6}The values of the gauge field function $H_2$ 
and the three Higgs field functions $\Phi_1$, $\Phi_3$ and $\Phi_4$ 
at the horizon $x_{\Delta}$ of the spherically symmetric
$n=1$ non-abelian black hole are shown as function of $x_{\Delta}$ for
the $\alpha\cdot C=0.5$, $C=\frac{1}{2}\sqrt{\frac{5}{3}}$,
and $\beta_2=\beta_3=1$.  }
\end{figure}
\end{fixy}

\begin{fixy}{-1}
\begin{figure}
\centering
\epsfysize=12cm
\mbox{\epsffile{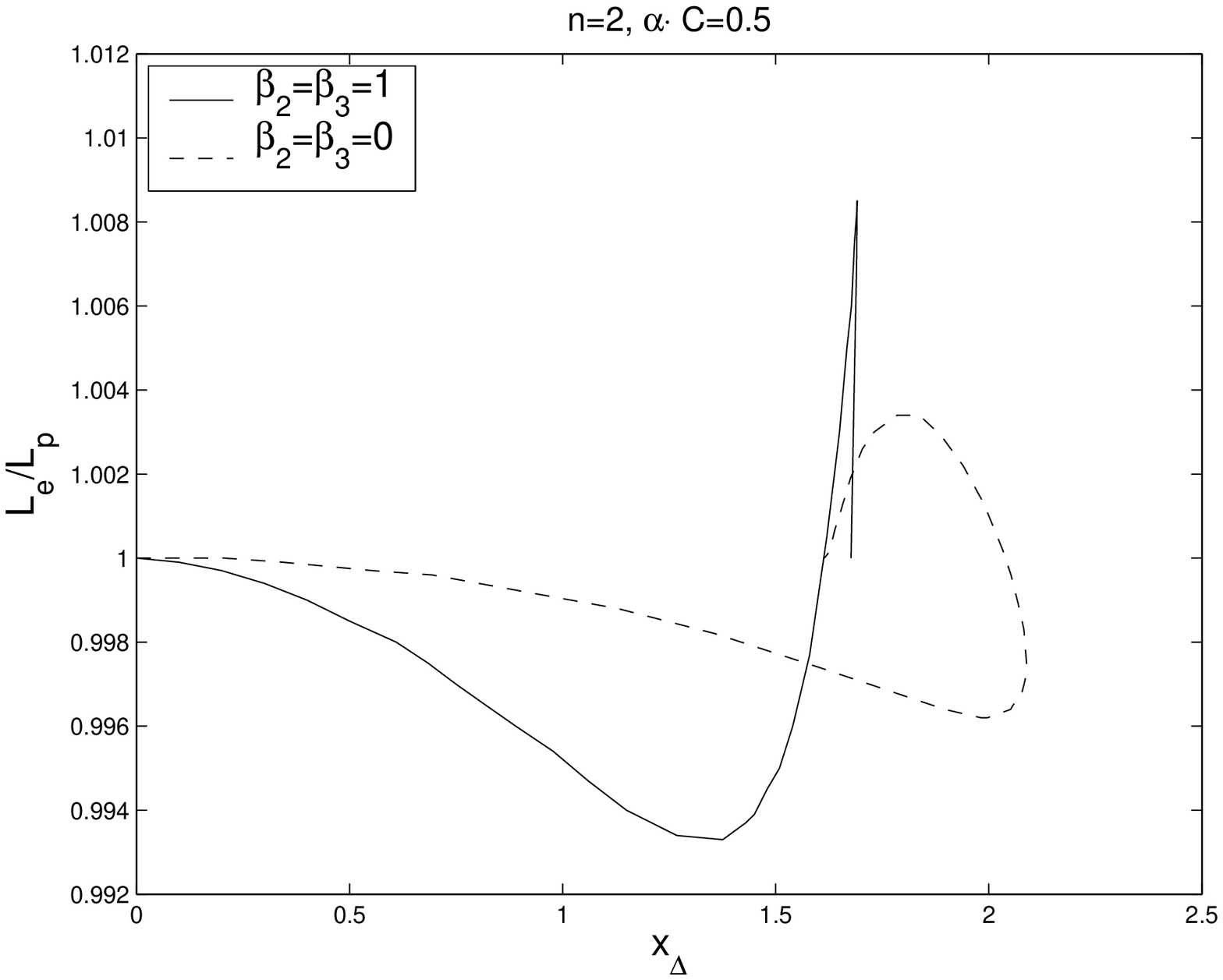}}
\caption{\label{Fig.7}The ratio of the circumference of the horizon
along the equator
$L_e$ and the circumference of the horizon along the poles $L_p$ is shown as function of
$x_{\Delta}$ for the axially symmetric $n=2$ non-abelian black holes 
for two values of $\beta_2=\beta_3$ and $\alpha\cdot C=0.5$,
$C=\frac{1}{2}\sqrt{\frac{5}{3}}$.  }
\end{figure}
\end{fixy}


\newpage
\begin{thebibliography}{ccc}
\bibitem{gg} H. Georgi and S. Glashow, Phys. Rev. Lett. {\bf 32}  (1974), 438. 
\bibitem{kt} E. Kolb and M. Turner, Addison Wesley, 1993.
\bibitem{ex} {\it see eg} : K. Hirata {\it et al.}, Phys. Lett. 
{\bf B220} (1989), 308;
R. Becker-Szendy {\it et al.}, Phys. Rev. {\bf D42} (1990), 2974;
M. Shiozawa {\it et al.}, Phys. Rev. Lett. {\bf 81} (1998), 3319. 
\bibitem{hp} G. 't Hooft, Nucl. Phys. {\bf B79} (1974), 276; 
A. Polyakov, JETP Lett.
{\bf 20} (1974), 194.
\bibitem{bogo} B. Bogomol'nyi, Sov. J. Nucl. Phys. {\bf 24} (1976), 449.
\bibitem{taubes} see e.g. A. Jaffe and C. Taubes, {\it Vortices and monopoles},
Birkh\"auser, 1980.
\bibitem{rr} 
C. Rebbi and P. Rossi, Phys. Rev. {\bf D22} (1980), 2010;
R. Ward, Commun. Math. Phys. {\bf 79} (1981), 317;
P. Forgacs, Z. Horvath and L. Palla, Phys. Lett. {\bf B99} (1981), 232;
E. Corrigan and P. Goddard, Commun. Math. Phys. {\bf 80} (1981), 575;
M. Prasad, Commun. Math. Phys. {\bf 80} (1981), 137;
M. K. Prasad and P. Rossi, Phys. Rev. {\bf D 24} (1981), 2182.n
\bibitem{ps} M. Prasad and C. Sommerfield, Phys. Rev. Lett. {\bf 35} (1975), 159.
\bibitem{manton} N. Manton, Nucl. Phys. {\bf B126} (1977), 525;
J. Goldberg, P. Yang, S. Park and K. Wali, Phys. Rev. {\bf D18} (1978), 542;
W. Nahm, Phys. Lett. {\bf B79} (1978), 426; Phys. Lett. {\bf B85} (1979), 373.
\bibitem{oyp} L. O'Raifeartaigh, S. Y. Yark and K. C. Wali, Phys. Rev. {\bf D20} (1979), 1941.
\bibitem{kkt} B. Kleihaus, J. Kunz and D. H. Tchrakian, Mod. Phys.
Lett. {\bf A13} (1998), 2523.  
\bibitem{hkk} B. Hartmann, B. Kleihaus and J. Kunz, Phys. Rev. Lett. {\bf 86} (2001), 1422;
Phys. Rev. {\bf D65} (2002), 0240027.
\bibitem{bh} Y. Brihaye and B. Hartmann, Phys. Lett. {\bf B528} (2002), 288; 
Phys. Lett. {\bf B534} (2002), 137;
B. Hartmann, Phys. Lett. {\bf B541} (2002), 368.
\bibitem{su3} E. Corrigan, D. Olive, D. Fairlie and J. Nuyts, 
Nucl. Phys. {\bf B106}
(1976), 475; T. Dereli and L. Swank, COO-3075-137 Yale print, (1976).
\bibitem{nb} Y. Brihaye and J. Nuyts, J. Math. Phys. {\bf 18} (1977), 2177.
\bibitem{tomaras} C. P. Dokos and T. N. Tomaras, Phys. Rev. {\bf D21} (1980), 2940;
D. Scott, Nucl. Phys. {\bf B171} (1980), 95.
\bibitem{eweinberg} E. Weinberg, Nucl. Phys. {\bf B167} (1980), 500.
\bibitem{newdata} It should be mentioned here, though, that in a recent
paper ({\it H. Adarkar, S. Dugad, M. Krishnaswamy, M. Menon and B. Sreekantan,
hep-ex/0008074}) results have been presented that suggest the lifetime of
the proton to be roughly $10^{31} years$, which would fit nicely
with an SU(5) GUT.
\bibitem{va} T. Vachaspati, Phys. Rev. Lett. {\bf 76} (1996), 188; H. Liu, T.
Vachaspati, Phys. Rev. {\bf D56} (1997), 1300.
\bibitem{meckes} M. Meckes, hep-th/0202001
\bibitem{va2} L. Pogosian and T. Vachaspati, Phys. Rev. {\bf D62} (2000), 105005.
\bibitem{marinov} E. B. Bogomol'nyi and M. S. Marinov, Sov. J. Nucl. Phys. {\bf 23} (1976), 357.
\bibitem{hkk2} B. Hartmann, B. Kleihaus and J. Kunz, Phys. Rev. {\bf D65} (2002), 024027.
\bibitem{gsu2} P. van Nieuwenhuizen, D. Wilkinson and M. Perry, Phys. Rev. {\bf 13} (1976), 778;
K. Lee, V. P. Nair and E. J. Weinberg, Phys. Rev. {\bf D45} (1992), 2751;
P. Breitenlohner, P. Forgacs, D. Maison, Nucl. Phys. {\bf B383} (1992), 357; 
Nucl. Phys. {\bf B442} (1995), 126;
P. C. Aichelburg and P. Bizon, Phys. Rev. {\bf D48} (1993), 607.
\bibitem{lw} A. Lue and E. J. Weinberg, Phys. Rev. {\bf D60} (1999), 084025;
Y. Brihaye, B. Hartmann and J. Kunz, Phys. Rev. {\bf D62} (2000), 044008. 
\bibitem{bp} Y. Brihaye and B. Piette, Phys. Rev. {\bf D64} (2001), 084010.
\end{thebibliography}
\end{document}